\documentclass[prb,aps,twocolumn,showpacs]{revtex4}
\usepackage{graphicx,latexsym}
\usepackage{dcolumn}
\usepackage{amsmath,amssymb,epsf,bm}
\begin{document}
\title{Spin-Hall current and spin polarization in an electrically biased SNS Josephson junction}
\author{A.~G. Mal'shukov$^{1}$, C.~S. Chu$^{2,3}$}
\affiliation{$^1$Institute of Spectroscopy, Russian Academy of
Sciences, 142190, Troitsk, Moscow oblast, Russia \\
$^2$Department of Electrophysics, National Chiao Tung University,
Hsinchu 30010, Taiwan \\
$^3$National Center for Theoretical Sciences, Physics Division,
Hsinchu 30043, Taiwan}
\begin{abstract}
Periodic in time spin-Hall current and spin polarization induced by
a dc electric bias has been calculated in a superconductor-normal
2DEG-superconductor (SNS) Josephson junction. We assumed that the
band energies of electrons in the normal system are splitted due to
Rashba spin-orbit coupling. The transport parameters have been
calculated within the diffusion approximation and using perturbation
expansion over a small  SN contact transparency. We found out that
in contrast to the stationary Josephson effect,  the spin-Hall
current does not turn to zero. Besides a direct proximity effect
caused by Cooper pair's transition into a triplet state, the spin
current and polarization are also driven by a periodic electric
field associated with the charge imbalance.
\end{abstract}
\pacs{72.25.Dc, 71.70.Ej, 73.40.Lq}

\maketitle
\section{Introduction}
The spin-Hall effect (SHE) is a fundamental physical phenomenon
where the spin-orbit interaction (SOI) shows up in electron
transport on a macroscopic level. The interplay of spin precession
caused by SOI and electron acceleration in the electric field gives
rise to a flux of the out-of-plane spin polarization flowing
perpendicular to the electric current.  Although this effect was
predicted long time ago \cite{Dyakonov}, it has been observed
experimentally only recently in semiconductors \cite{Kato,Wunderlich} and
metals \cite{Valenzuela}.  The nature of this phenomenon is now well
understood and studied  for  various systems (for a review see
\onlinecite{Engel}).  The spin-Hall effect is being considered as a
tool  for manipulating electron spins in perspective spintronic
applications. On the other hand, the spin polarization accumulated
due to SHE is subject to dissipative processes of spin relaxation
and diffusion. From this point of view, it is interesting to
consider SHE in superconductors, as well as in SNS junctions, where
the N-region is represented by a normal electron system with a
strong enough spin-orbit coupling. Also, in such systems SHE
provides an opportunity for a direct coupling of spin degrees of freedom to
superconducting quibits \cite{quibits,Schuster}

An important distinction of spin-Hall effects in superconducting and
normal systems  is that in the latter case this effect is determined
by spin dynamics of single particles, while in the former case major
role is played by interference of triplet and singlet Cooper pairs
\cite{MalshukovSNS}.  The triplet correlations, in their turn, are induced in the condensate wave function by SOI \cite{Gorkov}, and shows up as  admixture to the singlet state.  Therefore, the spin current and spin
accumulation caused by SHE are determined by a coherent macroscopic
state and do not dissipate. SHE has been considered in bulk
superconductors \cite{Contani} and SNS tunneling contacts
\cite{MalshukovSNS}. Also in such contacts an effect  reciprocal to
SHE was recently studied \cite{MalshukovISNS}.  These studies have
been restricted to the stationary transport. In the case of SNS
junctions this means that the Josephson electric current is driven
by a phase difference of superconducting order parameters  of  two
superconducting electrodes. If these electrodes have different
electric potentials, this current will periodically vary in time. One
would expect that, due to such a time dependence, the non-zero
spin-Hall current will be induced, while it is forbidden in the
stationary case by the time inversion symmetry. \cite{MalshukovSNS}
One more nonstationary effect is associated with an electric field caused by a dynamic electron-hole charge imbalance within the normal layer. Such
a periodic field will drive a flux of the spin polarization,  in a
way quite similar  to conventional SHE in normal systems.
\begin{figure}[bp]
\includegraphics[width=9cm]{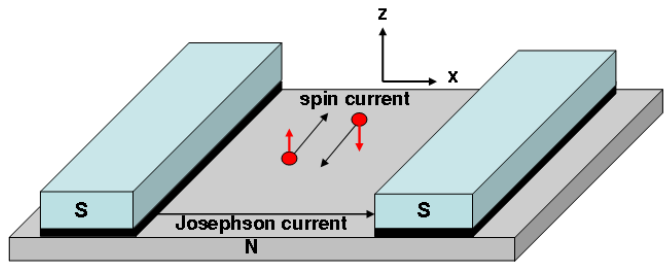}
\caption{(Colour online) An SNS Josephson junction. The ac Josephson
electric current flows in $x$-direction between two superconducting
electrodes (blue) through normal 2DEG (gray). Black layers show
tunneling barriers. In the presence of Rashba spin-orbit coupling
the ac spin current of $z$-polarized electrons in $y$-direction is
induced, with zero z-oriented spin density. A finite
spin polarization in y-direction (not shown) is also induced .}
\end{figure}

The goal of the present study is to extend the theory of
Ref.\onlinecite{MalshukovSNS} to the case of  the non-stationary
Josephson effect. We will calculate the spin-Hall current and spin
polarization created by a combined effects of the Josephson tunneling and
SOI. It  will be assumed that two singlet superconducting electrodes
are under the dc electric voltage $V$. The normal layer is contacted
to them through the low transparency tunneling barriers, as shown in
Fig.1. This layer is taken thin enough, so that electrons are
restricted to 2D motion, as, for example in a semiconductor quantum
well. The spin-orbit coupling  in the N-layer is given by the Rashba
interaction \cite{RashbaSOI} and we will ignore the spin-orbit
effects in scattering of electrons from impurities. At the same time
the spin-independent  scattering will be taken into account within
the Born approximation. The particle's mean free path $l$ will be
assumed smaller than all relevant parameters of length dimension,
except the Fermi wavelength $k_F$, which in the semiclassical
approximation is much smaller than $l$. Therefore, the electron
transport within the N-layer is dominated by diffusion.

The outline of this paper is as follows. The general expressions for
the spin-Hall current and spin density are derived in Sec. II. In
Sec. III some numerical results are presented and discussed. Finally, Appendix A presents some details of analytical calculations within the Keldysh formalizm.

\section{Basic equations}

Since we will focus on the basic characteristics of SHE, the
simplest approach will be employed within the lowest-order
perturbation theory with respect to transmission coefficients of
interface barriers. It should be noted, however, that higher-order
corrections to the electric Josephson current are not always small
\cite{Feigelman}, in particular in the range of temperatures $T$
larger than the Thouless energy $E_{Th}=D/L^2$, where $D$ is the
diffusion constant and $L$ is the distance between contacts. We will
assume the temperature or/and the transmission coefficient low
enough to avoid such a situation. The tunneling is presented by the
perturbation Hamiltonian
\begin{equation}\label{Hint}
\hat{H}_{int}=\sum_{k,k^{\prime},\sigma}t_{k,k^{\prime}}\hat{a}^{\dag}_{k^{\prime}\sigma}\tau_3 \hat{c}_{k\sigma} + h.c.\,,
\end{equation}
where $a_{k\sigma}$ and  $c_{k\sigma}$ are electron destruction
operators in the superconductor and N-layer, respectively, with $k$
denoting the wavevector and $\sigma$ the spin projection of the
particles. The transmission coefficient will be assumed a slow
varying function of wavevectors in the vicinity of the Fermi
surface.
Hamiltonian (\ref{Hint}) is written in the Nambu representation,
where destruction operators are defined as
\begin{equation}\label{Nambu}
\hat{c}_{k\sigma}=\left(\begin{array}{cc} c_{k\sigma} \\ c^{\dag}_{-k\bar{\sigma}}\end{array}\right),
\end{equation}
and $\tau_1, \tau_2, \tau_3 $ are the Pauli matrices in the Nambu
space. In its turn, the unperturbed Hamiltonian of  the normal layer
has the form
\begin{eqnarray}\label{H0}
\hat{H}_0&=&\sum_{k, \sigma,\sigma^{\prime}}
\hat{c}^{\dag}_{k\sigma}\left(\frac{\delta_{\sigma\sigma^{\prime}}\hat{\tau_3}}{2m^*}k^2
-\delta_{\sigma\sigma^{\prime}}\hat{\tau}_3
\mu+\bm{\sigma}_{\sigma\sigma^{\prime}}\cdot
\mathbf{h}_{\mathbf{k}}\right)\hat{c}_{k\sigma^{\prime}}+
\nonumber \\
&&\sum_{k,k^{\prime},\sigma}U_{k,k^{\prime}}\hat{c}^{\dag}_{k^{\prime}\sigma}\tau_3
\hat{c}_{k\sigma} + H_c \,,
\end{eqnarray}
where $\bm{\sigma}=(\sigma^x,\sigma^y,\sigma^z)$ is the Pauli spin
vector. The Rashba spin-orbit field $\mathbf{h}_{\mathbf{k}}$, which
is the odd function of $\mathbf{k}$, is given by \cite{RashbaSOI}
$\hat{h}^{x}_{\mathbf{k}}=\alpha k_y,
\hat{h}^{y}_{\mathbf{k}}=-\alpha k_x$. The random impurity
scattering potential is represented by its matrix elements
$U_{k,k^{\prime}}$. This scattering determines the elastic mean free
time $\tau$ of electrons. For a short-range scattering it is given
by $1/\tau=2\pi N_F \langle |U_{k,k^{\prime}}|^2
\rangle_{\text{imp}}=2\pi n_i N_F |U|^2$ , where $N_F$ is the state
density at the Fermi level, $n_i$ is the impurity concentration and
the subscript "imp" denotes averaging over impurity positions. The Hamiltonian $H_c$ represents the Coulomb interaction of electrons. It will be treated within the random phase approximation to take into account screening effects associated with the dynamic charge imbalance, while its contribution to the self-energy and electron-electron correlations will be ignored.

We assume that SNS contact is unbounded in the $y$-direction. Hence,
the Josephson current is in the $x$-direction, as shown in Fig. 1,
and the spin-Hall current  polarized parallel to the $z$-axis flows
in the $y$-direction and depends on the $x$-coordinate. In the
framework of the Keldysh formalism \cite{Landau} it can be written
as
\begin{equation}\label{Js}
J_s(x,t)=\frac{1}{4m^*}(\nabla_{y^{\prime}}-\nabla_y) \mathrm{Tr}[
\sigma^z\langle
G^K_{11}(t,\mathbf{r};t,\mathbf{r}^{\prime})\rangle_{\text{imp}}]|_{\mathbf{r}\rightarrow
\mathbf{r}^{\prime}} \,,
\end{equation}
where $G^K_{11}(t,\mathbf{r};t,\mathbf{r}^{\prime})$ is a
nondiagonal (Keldysh) component of the Green function, which is a
2$\times$2 matrix in the spin space, with subscript 11 denoting the
corresponding projection in the Nambu space. Besides the spin
current we will calculate also the spin polarization along the
$y$-axis. In normal systems such a spin polarization is due to the
electric spin orientation \cite{Edelstein}. It  is usually
associated with SHE and takes place also at stationary Josephson
tunneling conditions \cite{MalshukovSNS}. This polarization can be expressed as
\begin{equation}\label{Sy}
S_y(x,t)=-\frac{i}{2}\mathrm{Tr}[\sigma^y\langle
G^K_{11}(t,\mathbf{r};t,\mathbf{r})\rangle_{\text{imp}}]
\end{equation}

Expressions (\ref{Js}) and (\ref{Sy}) have to be expanded up to the
4-th order with respect to the transmittance. The relevant diagrams
are shown in Fig.2. Comparing them to a diagram representation of
the charge Josephson current \cite{Aslamazov} one can see that in
the latter case diagrams are much simpler. That is because
conservation of the Josephson current allows to reduce its
calculation to calculation of the charge time derivative in one of
the superconducting terminals. Unlike the Josephson current, the
spin current does not conserve and such a simplification is not
possible. An additional problem is caused by time dependence of the
charge transport. It results in an electric potential inside the
N-layer. Therefore, one should  take into account diagrams which
explicitly take into account the Coulomb screening there. These
diagrams are shown in Fig. 3. The calculations below will be
restricted to the low temperature $T$ and small enough DC voltage
$V$, both much less than the superconducting gap $\Delta$. In this
regime the electron transport through the contact will be dominated
by tunneling of Cooper pairs between two superconducting electrodes,
while a  dissipative transport due to electronic excitations will be
exponentially suppressed. Henceforth, the Green functions of
superconductor terminals are represented in Fig. 2 and Fig. 3 by
corresponding anomalous functions. More details on calculation of Feynman diagrams in Figs. 2 and 3 can be found in Appendix A. 
\begin{figure}[tp]
\includegraphics[width=4cm]{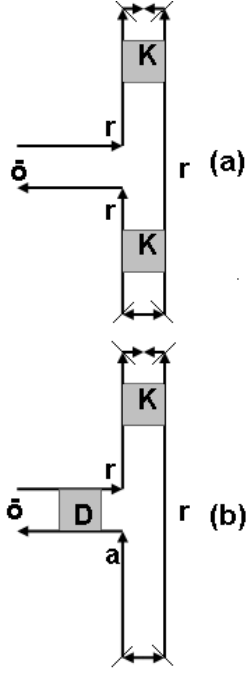}
\caption{(Color online) Examples of Feynman diagrams for
calculation of the spin current and spin polarization. The operator
$\bar{\mathbf{O}}$ denotes the spin current, or spin density
operators. Grey boxes are diffusion propagators. Boxes denoted by
"K" and "D" relate to Cooperon and diffuson, respectively. "r" and
"a" stand for retarded and advanced Green functions. Slashes denote
the tunneling amplitude}
\end{figure}

The main building blocks of diagrams in Fig. 2 and Fig. 3 are
unperturbed equilibrium Green functions averaged over impurity
positions.  For the normal layer these functions are determined by
Hamiltonian (\ref{H0}) and are represented by their retarded ($r$),
advanced ($a$) and Keldysh components
\begin{equation}\label{G0}
\hat{G}^{0r(a)}(\omega,\mathbf{k})=\left(\omega-\tau_3
E_{k}-\bm{\sigma}\cdot \mathbf{h}_{\mathbf{k}}\pm
i\Gamma\right)^{-1}\,,
\end{equation}
where $E_k=(k^2/2m^*)-\mu$  and $\Gamma=1/2\tau$,
\begin{equation}\label{G0K}
\hat{G}^{0K}(\omega,\mathbf{k})=\tanh\frac{\omega}{2k_BT}
\left(\hat{G}^{0r}(\omega,\mathbf{k})-\hat{G}^{0a}(\omega,\mathbf{k})\right)\,.
\end{equation}
Important entries in Fig. 2 are the propagators $\mathrm{D}$ and
$\mathrm{K}$ given by
\begin{eqnarray}\label{D}
&&D_{\alpha\beta\nu\mu}(\omega_1-\omega_2)= \nonumber \\
 &&n_i |U|^2 \langle
G_{\alpha\mu;11}^{r}(\mathbf{r},\mathbf{r}^{\prime},\omega_1)G_{\nu\beta;11}^{a}(\mathbf{r}^{\prime},\mathbf{r},\omega_2)\rangle_{\text{imp}}
\end{eqnarray}
and
\begin{eqnarray}\label{K}
&&K_{\alpha\beta\nu\mu}^{r(a)}(\omega_1+\omega_2)= \nonumber
\\&&n_i |U|^2 \langle
G_{\alpha\mu;11}^{r(a)}(\mathbf{r},\mathbf{r}^{\prime},\omega_1)G_{\nu\beta;22}^{r(a)}(\mathbf{r}^{\prime},\mathbf{r},\omega_2)\rangle_{\text{imp}}.
\end{eqnarray}
The conjugated functions $\mathrm{K}^{\dag}$ are defined by Eq.
(\ref{K}) with interchanged Nambu subscripts 11 and 22. Within the
semiclassical approximation $\mathrm{D}$ and $\mathrm{K}$ can be
represented by the ladder series. \cite{agd} At small frequencies
and large $|\mathbf{r}-\mathbf{r}^{\prime}| \gg l$  they satisfy a
diffusion equation and  are called "diffuson" and "Cooperon",
respectively. \cite{Altshuler} Due to the time inversion symmetry
these correlators are not independent. They can be expressed via
each other.
\begin{figure}[tp]
\includegraphics[width=5cm]{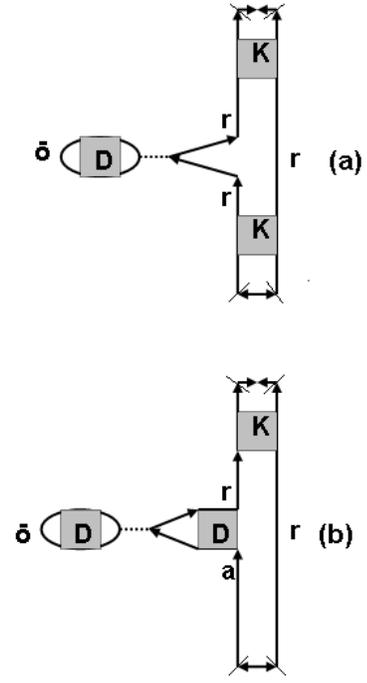}
\caption{(Color online) Examples of Feynman diagrams for
calculation of Coulomb screening effects on the spin current and
spin polarization (see Fig. 2 for details). Dashed line denotes the
screened Coulomb potential.}
\end{figure}
Depending on a combination of spin components, the diffuson and
Cooperon describe either spin, or particle (charge) diffusion.
Therefore, it is convenient to expand them in terms of Pauli
matrices, according to
\begin{equation}\label{D2}
D_{\alpha\beta\nu\mu}=\frac{1}{2}D_{ij}\sigma^i_{\alpha\beta}
\sigma^j_{\nu\mu}\,,
\end{equation}
where $i,j=x,y,z,0$ and $\sigma^0$ denotes the 2$\times$2 unity
matrix. Here and below a summation is assumed over the vector, or
spinor indexes entering twice into an expression. Cooperon components can be represented in a way similar to (\ref{D2}). The tensor components
$D_{ij}$ have a clear physical meaning. Namely, $D_{00}$ relates to
the particle diffusion, while various components with $i,j=x,y,z$
are associated with the spin diffusion. Mixed terms, for example
$D_{i0}$, are generally  not zero in the presence of SOI. As follows
from definitions (\ref{Nambu}) and (\ref{K}), $z$-components of
$\mathrm{K}$ are related to diffusion of singlet Cooper pairs,
because they involve antisymmetric combinations of "up" and "down"
spins in the particle-particle scattering process associated with
the propagator $\mathrm{K}$. Other components, 0,x,y are related to
triplet Cooperons. It is easy to see that the the 0-term gives a triplet
with a 0 spin projection onto the $z$-axis, while $x$ and $y$ components
are various combinations of $\pm 1$ triplets. A singlet-triplet
mixing is associated with nondiagonal correlators $K_{zm},K_{mz}$,
where $m=0,\pm 1$. As it will become clear below, the mixing terms
are proportional to the small parameter $\alpha/v_F$, where $v_F$ is
the Fermi velocity. Only linear in this parameter terms will be
taken into account in the following calculations of the spin-Hall
current and spin polarization.

Since $\mathrm{K}$ always enters together with the Green functions
of superconducting terminals, it is convenient to introduce the
pairing function
\begin{equation}\label{F}
F^{r(a)}_{\alpha\beta}(\mathbf{r},\omega)=\int d^2r^{\prime}K^{r(a)}_{\alpha\beta\nu\mu}(\mathbf{r},\mathbf{r}^{\prime},2\omega)\sigma_{\mu\nu}^z f^{r(a)}(\mathbf{r}^{\prime},\omega)\,,
\end{equation}
where $f^{r(a)}$ are determined by the anomalous superconductor
Green functions $\tilde{G}^{r(a)}_{12}$, as well as  by geometry of
contacts and their transmittance. Similarly, the conjugated
functions $F^{\dag r(a)}$ are defined through $\mathrm{K}^{\dag}$
and $f^{\dag }$. We assume that the SNS junction is symmetric, with
the electric potentials $\pm V/2$ applied to the left and right
electrodes, respectively. These potentials result in the time
dependent factors $\exp(\pm i\Omega (t+t^{\prime})/2)$ in
superconductor functions $\tilde{G}^0_{12}(t,t^{\prime})$ and
$\tilde{G}^0_{21}(t,t^{\prime})$, where $\Omega=\pm eV$.  In this
case, $f$ can be written in the form
\begin{equation}\label{f}
f^{r(a)}=-ia\frac{\Delta\left
[c(x-x_L)\delta_{\Omega,eV}+c(x-x_R)\delta_{\Omega,-eV}\right]}{\sqrt{(\omega
\pm i\delta)^2-\Delta^2}}\,,
\end{equation}
where $a$ can be expressed \cite{boundary}  through the
resistance $R_b$ of the SN interface, as  $a=1/4e^2 N_F R_b$ and
$c(x)$ is determined by a profile of the contact. For simplicity,
assuming that the distance between contacts $L=x_R-x_L$ is much
larger than the contact length, $c(x)$ will be approximated by the
delta-function. Since for our choice of  parameters $\Delta \gg
\omega,\Omega$, the retarded and advanced anomalous functions $f$
coincide, we will skip the labels $r,a$ below. The functions
$f^{\dag }$ are obtained from Eq. (\ref{f}) by the substitution
$L\rightarrow R$.

Let us introduce the vertex function
\begin{eqnarray}\label{j}
j_{l}(\Omega,\mathbf{q})=n_i|U|^2
\sum_{\mathbf{k}}G_{\alpha\beta;11}^{0a}(\omega,\mathbf{k})\frac{k_y\sigma^z_{\beta\gamma}}{m^*}\times
\nonumber \\
G^{0r}_{\gamma\rho;11}(\omega+\Omega,\mathbf{k}+\mathbf{q})
D_{\rho\alpha\nu\mu}(\Omega,\mathbf{q})\sigma^l_{\mu\nu}\,.
\end{eqnarray}
Further, according to the diagram representation in Fig. 2 and Fig.
3 the spin-Hall current can be written as
\begin{eqnarray}\label{Js2}
&&J_s(x,t)=\sum_{q,q^{\prime},\Omega=\pm eV}\int \frac{d\omega}{2\pi}e^{i(q+q^{\prime})x}e^{-2i\Omega t}\times \nonumber \\
&&[J_{1}+J_{2}-j_0(2\Omega,Q)\tilde{V}_Q\frac{i\Omega N_F}{\Gamma}
(J_{1\text{scr}}+J_{2\text{scr}})]\,,
\end{eqnarray}
where $Q=q+q^{\prime}$, while $J_{1}$ and $J_{2}$ are given by
diagrams (a) and (b) in Fig. 2, respectively. Other terms in the
integrand take into account Coulomb screening, as depicted in Fig.
3. $\tilde{V}_q$ denotes the screened Coulomb potential
\begin{equation}\label{tildeV}
\tilde{V}_q=\frac{V_q}{\epsilon(2\Omega,q)}\,.
\end{equation}
At small $\Omega \ll E_F$ and $q \ll k_F$ the dielectric function $\epsilon(2\Omega,q)$ is represented by the hydrodynamic expression (see e.g. Ref. \onlinecite{Altshuler}).
\begin{equation}\label{epsilon}
\epsilon(2\Omega,q)=1+V_q\frac{N_F}{\Gamma} Dq^2 D_{00}(2\Omega,q)\,,
\end{equation}
where the diffusion propagator $D_{00}(\omega,q)$ is given by Eq. (\ref{Dyy}). Taking into account that the two-dimensional Coulomb interaction $V_q=2\pi/\epsilon_0q$,  it follows from Eqs. (\ref{epsilon}) and (\ref{Dyy}) that at small $\Omega$ and $q$ the second term in Eq. (\ref{epsilon}) dominates. Retaining in $\epsilon$ only this term we arrive to
\begin{equation}\label{tildeV2}
\tilde{V}_q=\frac{V_q\Gamma}{\Gamma+V_qN_FDq^2D_{00}(2\Omega,q)}\simeq
\frac{\Gamma}{N_FDq^2D_{00}(2\Omega,q)}
\end{equation}
This approximation corresponds to a complete screening of charge within the length scale much larger than the screening length.

Using  the above definitions of Green functions and correlators,
various terms in the integrand of Eq. (\ref{Js2}) can be expressed
in the form
\begin{widetext}
\begin{eqnarray}\label{J1}
J_1=\frac{1}{2}b^{rrr}_{zij}F^r_{i,q}(\omega+\frac{\Omega}{2})
F^{\dag r}_{j,q^{\prime}}(\omega-\frac{\Omega}{2})
\tanh\frac{\omega-\Omega}{2k_BT} -\frac{1}{2}
b^{aaa}_{zij}F^a_{i,q}(\omega+\frac{\Omega}{2})  F^{\dag
a}_{j,q^{\prime}}(\omega-\frac{\Omega}{2})
\tanh\frac{\omega+\Omega}{2k_BT}\,,
\end{eqnarray}
\end{widetext}
where $F_{i,q}$ are spacial Fourier transforms of $x$-coordinate
dependent functions $F_{i}$ defined as
\begin{equation}\label{Fi}
F_{i}=\frac{1}{\sqrt 2}Tr[\sigma^i F]\,.
\end{equation}
and the coefficients $b_{lij}$ are given by
\begin{equation}\label{b}
b^{\text{abc}}_{lij}(\mathbf{q},\mathbf{q}^{\prime})=\sum_{\mathbf{k}}\frac{k_y}{m^*}
Tr[\Lambda^{\text{abc}}_{lij}(\mathbf{q},\mathbf{q}^{\prime})]
\end{equation}
with $\Lambda$ defined by the expression
\begin{eqnarray}\label{lambda}
&&\Lambda^{\text{abc}}_{lij}(\mathbf{q},\mathbf{q}^{\prime})=
-i\sum_{\mathbf{k}}G_{11}^{0\text{a}}(\omega-\Omega,\mathbf{k}-\mathbf{q}^{\prime})\times
\nonumber \\ &&\sigma_l
G^{0\text{b}}_{11}(\omega+\Omega,\mathbf{k}+\mathbf{q})
\sigma_iG^{0\text{c}}_{22}(\omega,\mathbf{k})\sigma_j \,.
\end{eqnarray}
Each of the symbols $\text{a,b,c}$ take the values $r$, or $a$,
while the subscripts $l,i,j$ ran through $0,x,y,z$. Introducing also
the coefficients
\begin{equation}\label{c}
c^{\text{abc}}_{lij}=\sum_{\mathbf{k}}Tr[\Lambda^{\text{abc}}_{lij}]\,,
\end{equation}
other terms in Eq.(\ref{Js2}) are expressed as
\begin{widetext}
\begin{eqnarray}\label{J2}
J_2=\frac{1}{4}j_{l}(2\Omega,\mathbf{q+q^{\prime}})c^{arr}_{liz}
F^r_{i,q}(\omega+\frac{\Omega}{2})
f^{\dag}_{q^{\prime}}(\omega-\frac{\Omega}{2})\left(\tanh\frac{\omega}{2k_BT}
-\tanh\frac{\omega-\Omega}{2k_BT}\right)+
\nonumber \\
\frac{1}{4}j_{l}(2\Omega,\mathbf{q+q^{\prime}})c^{ara}_{lzi}
f_q(\omega+\frac{\Omega}{2})
F^{\dag a}_{i,q^{\prime}}(\omega-\frac{\Omega}{2})\left(
\tanh\frac{\omega+\Omega}{2k_BT}-\tanh\frac{\omega}{2k_BT}\right)\,,
\end{eqnarray}

\begin{equation}\label{J1scr}
J_{1\text{scr}}=\frac{1}{2}c^{rrr}_{0ij}F^r_{i,q}(\omega+\frac{\Omega}{2})
F^{\dag r}_{j,q^{\prime}}(\omega-\frac{\Omega}{2})
\tanh\frac{\omega-\Omega}{2k_BT} -
\frac{1}{2}c^{aaa}_{0ij}F^a_{i,q}(\omega+\frac{\Omega}{2})  F^{\dag
a}_{j,q^{\prime}}(\omega-\frac{\Omega}{2})
\tanh\frac{\omega+\Omega}{2k_BT}
\end{equation}

\begin{eqnarray}\label{J2scr}
J_{2\text{scr}}=\frac{1}{2}D_{00}(2\Omega,\mathbf{q+q^{\prime}})c^{arr}_{0iz}F^r_{i,q}(\omega+\frac{\Omega}{2})
f^{\dag}_{q^{\prime}}(\omega-\frac{\Omega}{2})\left(\tanh\frac{\omega}{2k_BT}-
\tanh\frac{\omega-\Omega}{2k_BT}\right)+
\nonumber
\\\frac{1}{2}D_{00}(2\Omega,\mathbf{q+q^{\prime}})c^{ara}_{0zi}
f_q(\omega+\frac{\Omega}{2})
F^{\dag a}_{i,q^{\prime}}(\omega-\frac{\Omega}{2})\left(
\tanh\frac{\omega+\Omega}{2k_BT}-\tanh\frac{\omega}{2k_BT}\right)\,,
\end{eqnarray}
\end{widetext}

The spin density given by Eq. (\ref{Sy}) is calculated in a way
similar to the spin current. The same equation as Eq. (\ref{Js2})
can be used with the following changes: in Eq.(\ref{J1}) the factors
$c_{yij}$ from Eq.(\ref{c}) should be used instead of  $b_{zij}$; in
Eq. (\ref{J2}) one should substitute $2D_{yl}(2\Omega,q)$
instead of $j_{l}(2\Omega,q)$. Also, in Eq.
(\ref{Js2}) the vertex $j_0(2\Omega,q)$ should be substituted for
$2D_{y0}(2\Omega,q)$.

Before proceeding with further calculation of the spin-Hall current
and spin density, it is useful to discuss the physical meaning of
Eqs. (\ref{J1}), (\ref{J2}) and (\ref{J1scr})-(\ref{J2scr}). $J_1$
gives a "bulk" contribution to the spin current (spin density). It
is determined by diffusion  of Cooper pairs from the left and right
superconducting leads. Since for a chosen range of parameters the
distances from the leads are much larger than the coherence length
in the normal metal $\sqrt{D/\Delta}$, where $D$ is the diffusion
constant, the penetration depth of Cooper pairs into the normal
metal is determined by the diffusion length during the time much
larger than $\Delta^{-1}$. Thus, the characteristic diffusion time
of singlet pairs is of the order of min[$(k_B T)^{-1}, (eV)^{-1}$],
while in the case of triplets the spin relaxation time comes into
play, if it is shorter than the diffusion time of singlets.
Therefore, if the spin relaxation time is shorter than the Thouless
time $E_{Th}^{-1}=L^2/D$, the triplet components $F_i$ ($i=0,x,y$)
of the pairing function will be localized relatively close to the
leads. At the same time, at $E_{Th} \gtrsim$ max[$k_B T,eV$] the
singlets $F_z$ can propagate through entire junction. As it will be
shown in the next section, $J_1$ is represented by a combination of
a pure singlet term of the form $F_z F^{\dag}_z$ and singlet-triplet
interference contributions, like $F_z F^{\dag}_x$. The former can
penetrate over large distances, independent on the magnitude of the
spin-relaxation rate associated with the spin-orbit coupling.
Therefore, at low enough $T$ and $V$ the spin current represented by
$J_1$ in Eq. (\ref{Js2}) can be observed far from the contacts. At
the same time, the spatial distribution of the current given by
$J_2$ is determined by the spin density created by one-particle spin
diffusion near one of the contacts. This diffusion is represented by
the diffusion propagator $D$ in Fig. 2b. The diffusion in this case
is restricted by the spin relaxation length. If this length is less
than $L$ the corresponding spin current (spin density) will be
distributed relatively close to contacts. So, it is of the "surface"
type. The remaining screening terms in Eq. (\ref{Js2}) are
determined by the long-range Coulomb interaction. Therefore, their
contribution will be of the "bulk" type.

\section{Results and discussion}

In this section we will calculate the functions and coefficients entering into
general expressions Eqs.(\ref{Js2})-(\ref{J2scr}) and present numerical results for spin transport parameters.

The matrix $D_{ij}$ can be found from the diffusion equation. In the
SHE regime this equation has been derived in a number of works.
\cite{Misch} The Cooperon $K_{ij}$ can, in its turn, be  expressed
through $D_{ij}$. Since the latter depends only on the
$x$-coordinate, the diffusion drift of particles occurs in the
$x$-direction. Hence, the effective "magnetic" field
$\mathbf{h}_{\mathbf{k}}$ induced by the Rashba interaction is
directed parallel to the $y$-axis. Therefore, electron spins precess
in the $zx$ plane. This means that one has coupled equations for
$D_{zj}$ and $D_{xj}$ , while $D_{yj}$-components  are decoupled
from them. They, however, stay coupled to the "charge" diffuson
$D_{0j}$ through the weak spin-charge coupling . For example, the
mixed function $D_{y0}(\omega,x)$ satisfies the equation
\begin{equation}\label{diffeq}
-i\omega D_{y0} -D\nabla_x^2 D_{y0}+\Gamma_{so}D_{y0}-2\Gamma\chi
\nabla_x D_{00}=0 \,,
\end{equation}
where $\chi=-\alpha \Gamma_{so}/4\Gamma^2$ and the D'ykonov-Perel' \cite{dp} spin relaxation rate is $\Gamma_{so}=\alpha^2 k_F^2/\Gamma$.
The spin-charge coupling $\chi$ will be taken into account in the lowest order perturbation expansion. Hence, after Fourier transformation Eq.(\ref{diffeq}) gives
\begin{equation}\label{Dy0}
D_{y0}=iq\chi D_{yy}D_{00}\,,
\end{equation}
where
\begin{equation}\label{Dyy}
D_{yy}=\frac{2\Gamma}{-i\omega
+Dq^2+\Gamma_{so}}\,\,;\,\,D_{00}=\frac{2\Gamma}{-i\omega +Dq^2}\,.
\end{equation}
When expressed through $D_{ij}$, the corresponding Cooperon components are
\begin{equation}\label{Kzx}
K^r_{xz}(\omega,q)=-K^a_{xz}(-\omega,q)=-iD_{y0}(\omega,q)
\end{equation}
and $K_{xz}=K^{\dag}_{xz}$. At the same time
\begin{equation}\label{Kzz}
K^r_{zz}(\omega,q)=K^a_{zz}(-\omega,q)=-D_{00}(\omega,q)
\end{equation}

Further,  keeping only the leading terms with respect to small
parameters  $\Omega\tau, \omega\tau, q\tau$ and $h_k\tau$, from Eqs.
(\ref{b}), (\ref{lambda}) and (\ref{c}) one can easy calculate the
factors $b$ and $c$. The coefficients $c$ are given by
\begin{eqnarray}
&&c^{rrr}_{yxz}=c^{aaa}_{yzx}=c^{arr}_{yzx}=c^{ara}_{yxz}=-i\frac{\pi
N_F}{\Gamma^2} \label{c1}\\
&&c^{arr}_{0zz}=c^{ara}_{0zz}=-c^{rrr}_{0zz}=c^{aaa}_{0zz}=-\frac{\pi N_F}{\Gamma^2}\,.
\end{eqnarray}
In Eq. (\ref{c1}) these coefficients change their signs with each
permutation of lowercase indexes. The factors $b$, in their turn,
are defined by
\begin{eqnarray}
&&b^{rrr}_{zxz}=b^{rrr}_{zzx}=b^{aaa}_{zxz}=b^{aaa}_{zzx}=-i\frac{\alpha k_F^2\pi
N_F}{2m^*\Gamma^3} \label{b1}\\
&&b^{rrr}_{zzz}=-b^{aaa}_{zzz}=i(q+q^{\prime})\frac{\alpha^2
k_F^2\pi N_F}{4m^*\Gamma^4} \label{b2}
\end{eqnarray}
and the vertex functions $j_i$ calculated from Eqs. (\ref{j}) and
(\ref{Dy0}) are given by
\begin{eqnarray}\label{j0y}
j_0(2\Omega,Q)&=&iQ\frac{\alpha^2
k_F^2}{2m^*\Gamma^2}\left(1-\frac{\Gamma_{so}}{2\Gamma}D_{yy}(2\Omega,Q)\right)\times
\nonumber  \\
&&D_{00}(2\Omega,Q)\,,\\
j_y(2\Omega,Q)&=&\frac{\alpha k_F^2}{m^*\Gamma}D_{yy}(2\Omega,Q)\label{j0y2}\,.
\end{eqnarray}

\begin{figure}[tp]
\includegraphics[width=8cm]{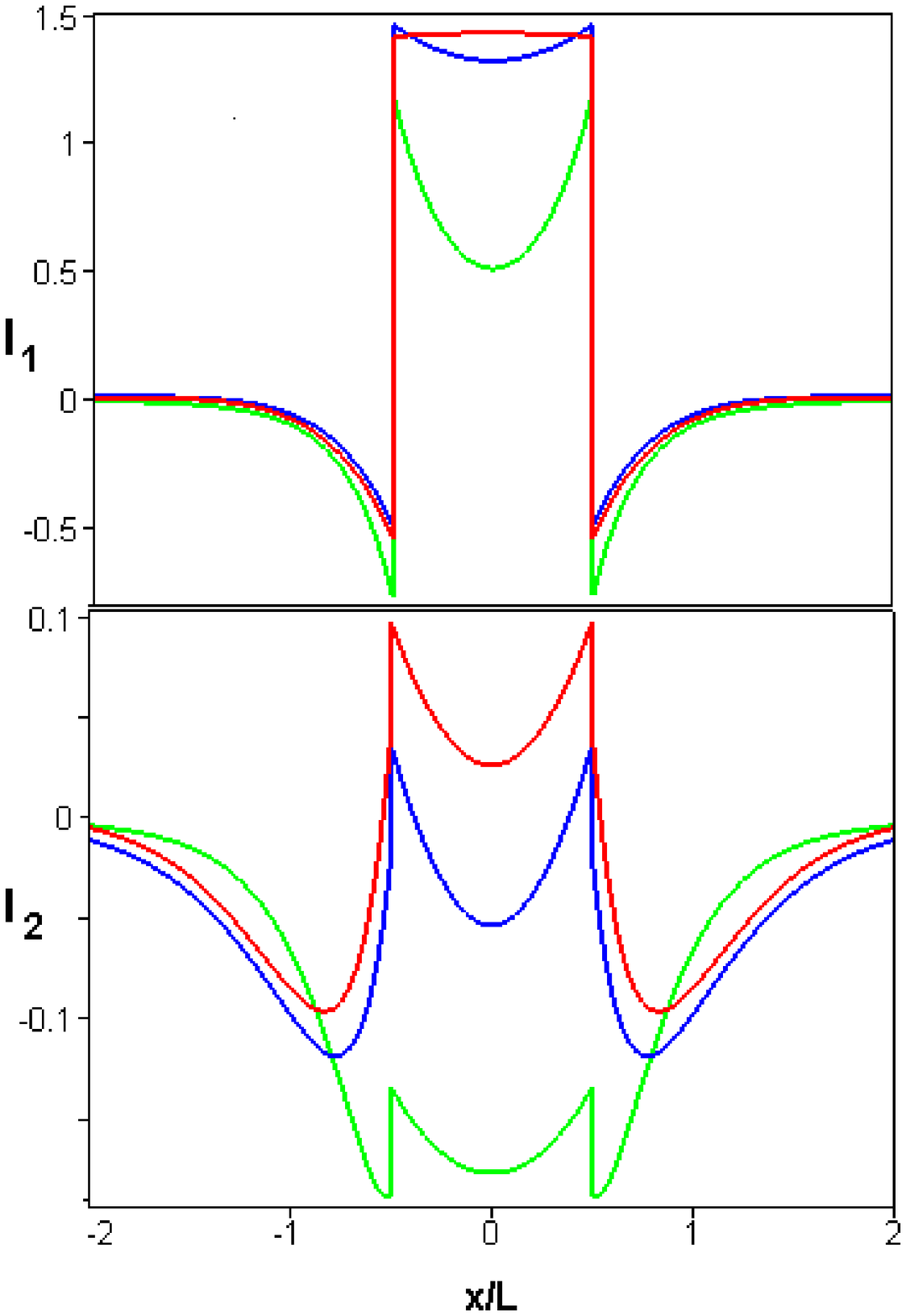}
\caption{(Color online) Coordinate dependence of two phase shifted
components of the spin-Hall current, as defined by Eqs.
(\ref{Jsnorma})  and (\ref{I12}). $\Gamma_{so}/E_{Th}$=0.1 (red), 1
(blue),and 10 (green). The curves are calculated at $\pi k_B
T/E_{Th}=0.5$ and $eV/E_{Th}=1$}.
\end{figure}
\begin{figure}[tp]
\includegraphics[width=8cm]{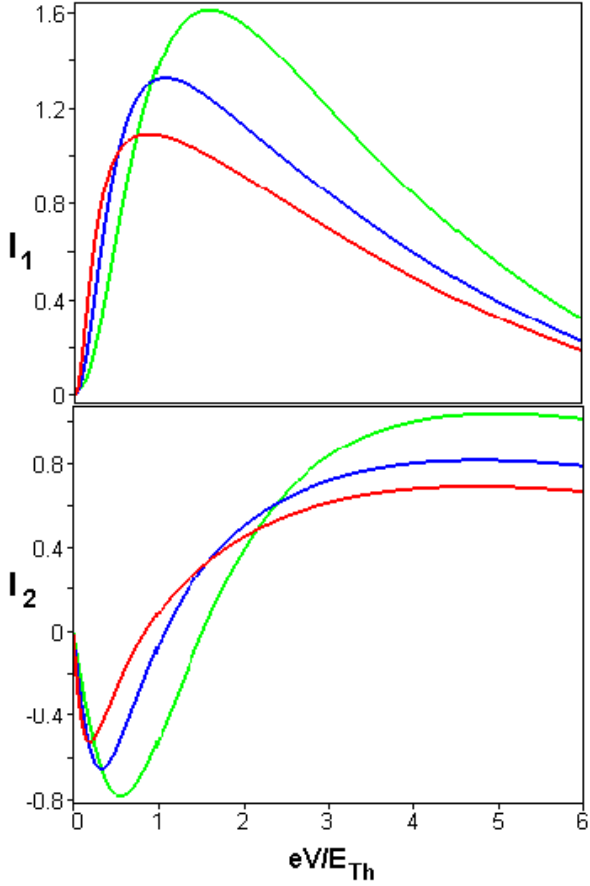}
\caption{(Color online) Spin-Hall current as a function of the bias
voltage at $\Gamma_{so}/E_{Th}=1$ and $x=0$. $\pi k_B T/E_{Th}$=0.25
(red), 0.5 (blue), and 1 (green).}
\end{figure}
A following important property of $J_1, J_2$ and
$J_{1\text{scr}},J_{2\text{scr}}$ calculated with functions and
coefficients defined by Eqs. (\ref{Dy0})-(\ref{j0y2}) takes place:
all these partial contributions to the spin-Hall current, after
initial increasing with the spin-orbit coupling $\alpha$, saturate
when $\alpha \rightarrow \infty$. This behavior is already seen in
$j_0(2\Omega,Q)$. Indeed, as follows from Eqs. (\ref{j0y}) and
(\ref{Dyy}), due to cancellation at $\Omega \rightarrow 0$ and $Q
\rightarrow 0$ of two terms in brackets of (\ref{j0y}) this function
becomes constant at large $\Gamma_{so} \sim \alpha^2$. It can be
checked that the same combination as in brackets of Eq.(\ref{j0y})
enters into all terms contributing to the spin-Hall current. This
sort of cancellation takes place also in normal systems and is
inherent to all linear in $k$ SOI couplings. In normal systems it
results in the vanishing stationary spin-Hall effect \cite{Engel}.
Indeed, there the spin current is driven by the stationary electric
field which, due to charge screening, at the small screening length
is homogeneous in samples of simple geometries. That guarantees
$Q=0$ and, hence, the vanishing spin-Hall current. In contrast, in
the considered here case of a nonstationary and inhomogeneous
electron transport the spin-Hall current remains finite. The reason
is that it is determined by the superconducting proximity effect, as
it is discussed in the end of Section II. Consequently, a finite
penetration range of Cooper pairs results in the finite $Q$.
Moreover, there are also the "surface" terms, like the $J_2$-term,
localized  near superconducting leads within the spin relaxation
length $\sqrt{D/\Gamma_{so}}$. For these terms  $DQ^2 \sim
\Gamma_{so}$. Therefore, they are not expected to saturate with
larger SOI.

\begin{figure}[tp]
\includegraphics[width=8cm]{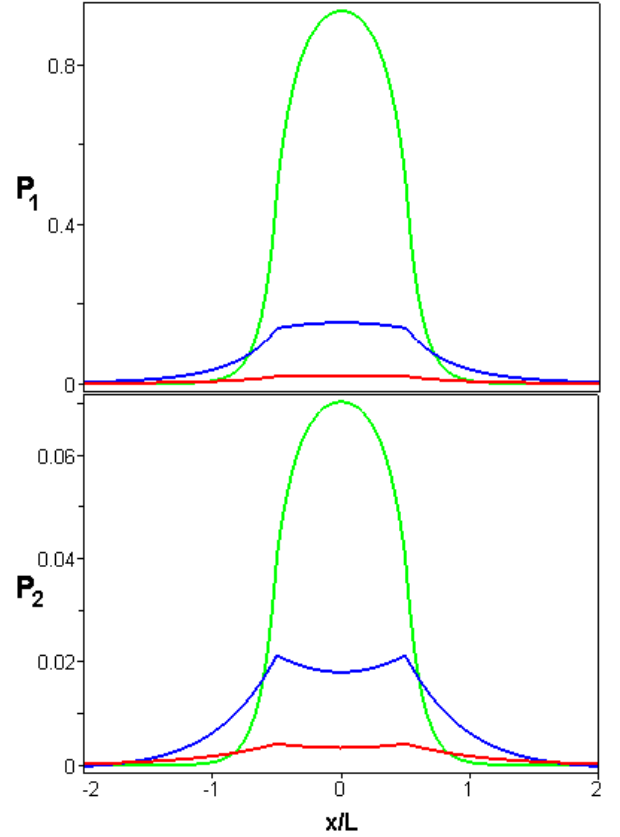}
\caption{(Color online) Coordinate dependence of two phase shifted
components of the spin polarization, as defined by Eqs. (\ref{P})
and (\ref{P12}). $\Gamma_{so}/E_{Th}$=0.1 (red), 1 (blue),and 10
(green). The curves are calculated at $\pi k_B T/E_{Th}=0.5$ and
$eV/E_{Th}=1$}.
\end{figure}
\begin{figure}[tp]
\includegraphics[width=8cm]{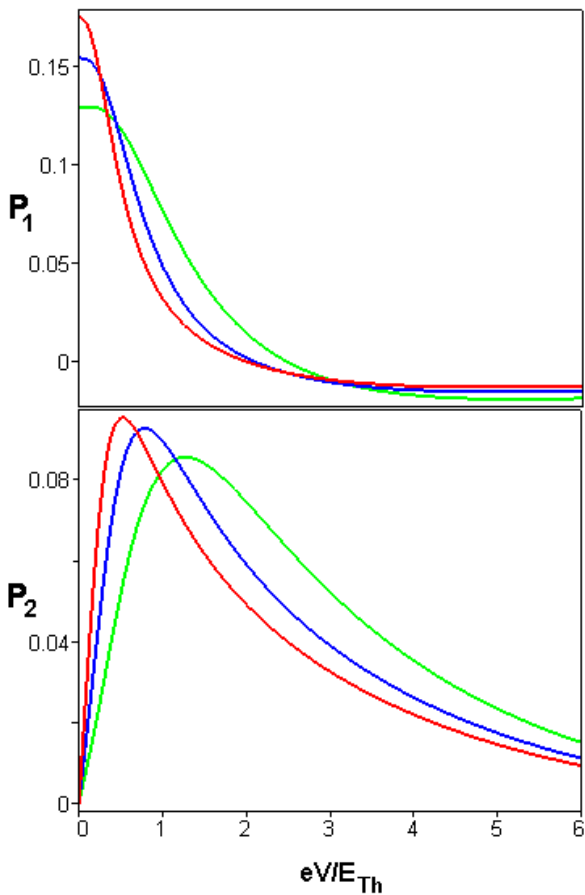}
\caption{(Color online) Spin polarization as a function of the bias
voltage at $\Gamma_{so}/E_{Th}=1$ and $x=0$. $\pi k_B T/E_{Th}$=1
(green), 0.5 (blue), and 0.1 (red).}
\end{figure}

We will normalize the spin-Hall current density according to
\begin{equation}\label{Jsnorma}
J_s=\frac{J_c}{e}\frac{m^* \alpha^2}{2\Gamma}I \,,
\end{equation}
where $J_c$ is the critical Josephson current density defined by the
sum over Matsubara frequencies $\omega=\pi k_B T (2n+1)$ as
\cite{Aslamazov}
\begin{equation}\label{Jc}
J_{c}=\frac{e4\pi N_F k_BT a^2}{\Gamma}
\sum_{\omega \geqslant 0}\frac{|\Delta|^2}{|\Delta|^2+\omega^2}D_{00}(2i\omega,L)\,,
\end{equation}
The diffuson in this equation is obtained as a Fourier transform
from (\ref{Dyy}) and has the form $D_{00}(2i\omega,L)=\Gamma
\exp(-\kappa L)/D\kappa$, with $\kappa=\sqrt{2\omega/D}$. The so
defined dimensionless factor $I$ is of the order of  1. We note that
although $\alpha$ can be quite large in narrow gap semiconductors
\cite{Nitta}, as well as in some metallic systems \cite{Ast}, the
parameter $m^* \alpha^2 /\Gamma$ in Eq. \ref{Jsnorma} is small,
because the diffusion approximation requires $k_F \alpha \ll
\Gamma$. 

The spin density is normalized as
\begin{equation}\label{P}
S_y=S_{y0}P \,,
\end{equation}
where $S_{y0}$ is the spin polarization induced by the critical
Josephson current in the stationary case. This polarization is given by
\cite{MalshukovSNS}
\begin{equation}\label{Sy0}
S_{y0}=\frac{\alpha \tau }{2e D}J_c \,,
\end{equation}
Since, according to (\ref{f}), integrand in (\ref{Js2}) contains
terms proportional to $\delta_{\Omega, \pm eV}$, the normalized spin
current and spin polarization can in general be represented as
\begin{eqnarray}\label{I12}
I&=&-I_1\sin(2eV t) + I_2\cos(2eV t) \\
P&=&P_1\sin(2eV t)- P_2 \cos(2eV t)\label{P12} \,.
\end{eqnarray}

The spatial distribution of $I_1$ and $I_2$ is shown in Fig. 4 at
different values of the spin-orbit couplings. It is seen that the
magnitude and direction of the spin-Hall current vary fast in the
region of contacts. We recall in this connection that the contacts are assumed
relatively narrow in Fig. 1 and are approximated by point-like
sources placed at $x=\pm L/2$.  At some moments of time the normalized spin
current (\ref{P12}) changes its sign also as a function of $\Gamma_{so}$, as one
can see from comparison of $I_2$ curves at $\Gamma_{so}=E_{Th}$ and
$\Gamma_{so}=10 E_{Th}$. Such changes are associated with discussed
above cancellation  of various contributions to the spin current at
$eV$ and $k_B T \ll \Gamma_{so}$.

It have been pointed out that there are two types of terms, "bulk"
and "surface" ones. At large $\Gamma_{so}$ the latter are localized
near contacts, while the former penetrate deeper between and outside
contacts. These qualitative features  are clearly seen in Fig. 4. As
expected, due to the spin-current saturation, the bulk contribution to the normalized current decreases with larger SOI. Indeed,
$I_1$ is reduced in the middle of the junction at
$\Gamma_{so}=10E_{Th}$ , while smaller changes are seen just near
the contacts. At the same time,  such a reduction is not so fast, as
it was expected at first sight. At least, even at $\Gamma_{so}=10
E_{Th}$ there is no considerable reduction of $I_2$ at $x=0$.

It is important to note that the spin current is not zero at
$|x|>1/2$, while the Josephson current is absent there. In this
region the former is driven by the time dependent potential,
associated with the charge imbalance, rather than by a direct
conversion of singlet Cooper pairs to triplet ones. In this spatial
region the spin current is contributed by both "surface" and "bulk"
terms. In Fig.4 it extends outside the junction over the range $\sim
L$, because the both characteristic lengths $\sqrt{D/k_B T}$ and
$\sqrt{D/eV}$  are taken of the order of $L$.

The voltage dependence of the spin-Hall current at $x=0$ and various
temperatures is presented in Fig. 5. We note that both phase shifted
components change sign at some voltages. At the same time, the
behavior at the small bias  is given by $I_1 \sim V^2$ and $I_2 \sim
V$, as can also be checked analytically.

The coordinate dependence of the normalized spin polarization is
shown in Fig. 6. In contrast to the considered previously stationary
case \cite{MalshukovSNS}, it is finite in the region $|x|>1/2$.
Similar to the spin-Hall current, such a behavior can be explained
by the charge imbalance effect. This effect becomes weaker at larger
$\Gamma_{so}$. Indeed, at  $\Gamma_{so}=10 E_{Th}$ the magnitude of
$P$ considerably increases inside the junction, while it becomes
smaller outside it at $|x|>1/2$.  This is opposite to the spin
current trend observed in Fig.4. The reason is that the slow
varying "bulk" terms are not suppressed in $P$ at larger SOI,
because in contrast to the spin current, there is no cancellation of
the spin polarization at $DQ^2$ and $\Omega \ll \Gamma_{so}$.

A variation of the polarization with $V$ is presented in Fig. 7 for
$\Gamma_{so}=E_{Th}$ . $P_2$ linearly turns to 0 at $V
\rightarrow 0$, while  $P_1$ reaches its maximum there. Our
numerical results show also that the magnitude of $P_1$ at $V=0$
increases with $\Gamma_{so} \rightarrow \infty$, reaching 1, that
can also be checked  analytically.  At the same time, the
spatial dependence of $P_1$ takes the form of a step function, such
that $P_1=1$ when $|x|<0.5 L$ and $P_1=0$ at $|x|>0.5 L$. Hence,  as
expected, in this limiting case the spin polarization coincides with polarization calculated in the regime of the stationary Josephson effect \cite{MalshukovSNS}, after
substitution in (\ref{P12}) of the phase factor $\sin(2eVt)$ by
$\sin \phi$, where $\phi$ is the phase difference between superconductors.

In the considered here case the spin-Hall current carries the
z-oriented time dependent spin polarization in the transverse (y)
direction. Hence, the corresponding time dependent spin density can
be accumulated at some distance near flanks of the junction, while
in its bulk the y-polarization is finite between
and outside contacts. The spin polarization might be detected by
various methods employed in the case of the ordinary SHE
\cite{Kato,Valenzuela}. Quite efficient is an all-electric method
used in Ref. \onlinecite{Valenzuela}. In this method a
nonequilibrium spin polarization in the normal metal diffuses into
an adjacent ferromagnet. On the other hand, it is well known
\cite{Johnson}, that the spin flux through a ferromagnet-normal
metal interface induces a voltage difference across the interface,
that can be measured. If we will try, however,  to extend this
method to the superconducting transport, we will face a problem of
evaluating this voltage. It is well known how to calculate it in the
case of a nonequilibrium flux of single particle spins. Much less,
however, is known how to do this in our case, when triplet Cooper
pairs contribute to this flux. Therefore, additional studies are
necessary for a  more complicated system than considered here.

This work has been supported by Taiwan NSC (Contract No.
96-2112-M-009-0038-MY3) and  MOE-ATU grant.

\appendix
\section{Derivation of basic equations}

In this Abstract some explanations will be done of how Eqs. (\ref{Js2}) and (\ref{J1})-(\ref{J2scr}) have been derived within Keldysh diagrammatic technique. We start
from definitions of Green functions entering into perturbation
expansions. Each of the Green functions is the matrix in the Keldysh
space \cite{Landau}:
\begin{equation}\label{Keldysh}
    G=\left(
      \begin{array}{ccc}
        \hat{G}^r  & \hat{G}^K \\
        0  & \hat{G}^a \\

      \end{array}
    \right)
\end{equation}
The elements of this matrix are 2$\times$2 matrices in the Nambu
space and 2$\times$2 matrices in the spin space. Besides, they
depend on two time arguments and two wavevectors $\mathbf{k}$ and
$\mathbf{k}^{\prime}$. These vectors are not equal, because there is
no momentum conservation in the presence of impurity scattering. At the same time, the
functions averaged over impurity positions become diagonal in the
momentum space. Considering $G$ as matrices in $k$-space and also
the tunneling amplitudes $t_{k,k^{\prime}}$ in Eq. (\ref{Hint}) as
elements of the matrix $\hat{t}$, one can write the expression for
the spin current (\ref{Js}) in the fourth perturbation order with
respect to the tunneling Hamiltonian (\ref{Hint}) in the form
\begin{widetext}
\begin{equation}\label{JsA}
    J_s(x,t)=\frac{1}{4}\int \prod_{i}dt_i\mathrm{Tr}[
\sigma^z\langle
\frac{\hat{k}}{m^*}\hat{G}_{11}(t-t_1)\hat{t}\tilde{\hat{G}}_{12}(t_1,t_2)\hat{t}
\hat{G}_{22}(t_2-t_3)\hat{t}\tilde{\hat{G}}_{21}(t_3,t_4)
\hat{t}\hat{G}_{11}(t_4-t)\rangle_{\text{imp}}]^K \,,
\end{equation}
\end{widetext}
where $\tilde{\hat{G}}$ and $\hat{G}$ denote the Green function of the
superconductor and the normal metal, respectively. The operator
$\hat{k}=k_y \delta_{\mathbf{k},\mathbf{k}^{\prime}}$. We explicitly
wrote the Nambu labels of functions, so that only the trace over
spin and momentum variables must be taken in Eq. (\ref{JsA}). Since
only the Josephson tunneling is considered, in the perturbation
expansion we take into account only anomalous $\tilde{\hat{G}}_{12}$ and
$\tilde{\hat{G}}_{21}$ functions of superconducting leads, neglecting thus
the usual stationary one-particle tunneling. These functions can be
written in the form
\begin{equation}\label{phase}
\tilde{\hat{G}}_{12}(t,t^{\prime})=\hat{\cal{F}}(t-t^{\prime})\exp(\pm
ieV(t+t^{\prime})/2)\,,
\end{equation}
where the signs "-" and "+" relate to the left and right
superconducting leads, respectively. $\tilde{\hat{G}}_{21}$ is obtained
from this equation with the substitution $\cal{F} \rightarrow \cal{F}^{\dag}$
and $V \rightarrow -V$. After Fourier transform of Green functions in Eq. (\ref{JsA}), the time dependent exponential factors in $\tilde{\hat{G}}_{12}(t,t^{\prime})$ and  $\tilde{\hat{G}}_{21}(t,t^{\prime})$ give the factors $\exp(\pm 2ieVt)$  in Eq. (\ref{Js2}). It is important that the functions $\cal{F}$ and
$\cal{F}^{\dag}$ in Eq. (\ref{JsA}) belong to different superconducting
leads. Therefore, they are independently averaged over impurity
positions and, hence, are diagonal in the momentum space. In the spin space they are proportional to the Pauli matrix $\sigma_z$, because, as it is discussed in Section II, these functions are associated with the singlet Cooper pairing. 

It is easy to see that the Keldysh component of a product $ABCD...$
of matrices having the triangular form (\ref{Keldysh}) can be
written as the sum of products: $(A^K B^a C^a D^a...)+(A^r B^K C^a
D^a...)+(A^r B^r C^K D^a...)+...$. In all these products the Keldysh
function enters only once, while retarded and advanced functions are
placed on the left and on the right from it, respectively. The time
Fourier expansions of  thermally equilibrium  Keldysh functions $G$
and $\cal{F}$ can be expressed in terms of retarded and advanced functions
as
\begin{equation}\label{Keldysh2}
    \hat{G}^K(\omega)= \left(\hat{G}^r(\omega)- \hat{G}^a(\omega)\right)
    \tanh\frac{\omega}{2k_B T}\,.
\end{equation}
We thus will apply the above expressions to the Keldysh component of
the product in Eq.(\ref{JsA}). This equation can be further
simplified taking into account that $k_BT$, $eV \ll \Delta$ and
distances of interest $\gg \sqrt{D/\Delta}$. Therefore, the main
contribution to Eq.(\ref{JsA}) is given by small frequencies $\omega
\ll \Delta$. Since the difference of retarded and advanced anomalous
functions $\cal{F}$ in Eq.(\ref{Keldysh2}) gives an expression
proportional to $\delta\left(E^2_k+\Delta^2 - \omega^2\right)$, the
corresponding Keldysh component can be neglected for small $\omega$.

The next step is averaging over disorder. From Eqs. (\ref{JsA}),
(\ref{Keldysh2}), and taking into account that $\hat{\tilde{G}} \sim
\sigma_z$ we get the following products to be averaged
$\hat{G}^r_{11}\sigma_z\hat{G}^r_{22}\sigma_z\hat{G}^r_{11}$,
$\hat{G}^a_{11}\sigma_z\hat{G}^a_{22}\sigma_z\hat{G}^a_{11}$,
$\hat{G}^r_{11}\sigma_z\hat{G}^a_{22}\sigma_z\hat{G}^a_{11}$, and
$\hat{G}^r_{11}\sigma_z\hat{G}^r_{22}\sigma_z\hat{G}^a_{11}$. In its
turn, each of the Green functions is a sum of products
$G_{k}U_{k,k_{1}}G_{k_{1}}U_{k_{1},k_{2}}...$, where $G_{k}$ are
diagonal in $\mathbf{k}$ unperturbed functions and the number of
impurity scattering amplitudes $U$ in each product is equal to the
corresponding perturbation order. Calculation of such averages is
described in many textbooks (for example, see
\onlinecite{Altshuler}). Briefly, within the Born approximation,
assuming random impurity positions the averages of
$U_{k,k_{\prime}}$ products decouples into pair averages. In this
way each pair enters as an effective two-particle interaction
carrying a zero frequency. So, the average of a Green function can
be expressed through the self-energy. In Eq. \ref{G0} this self-energy
is given by $i\Gamma$. When a product of Green functions is
averaged, a considerable simplification takes place in the
semiclassical approximation when $\Gamma \ll E_F$. In this case a
special class of the so called "ladder" diagrams dominates in the
perturbation expansion over disorder. Let us consider, for example
the average $\langle \hat{G}^r_{11}\sigma_z \hat{G}^r_{22}\sigma_z
\hat{G}^r_{11}\rangle_{\text{imp}}$. Combining in pairs $U$-s in
$G^r_{22}$ with its neighbors on the left and on the right we obtain
two ladder series. There are "Cooperons", according to definition
(\ref{K}). They are shown as gray boxes in Fig. 2a. At the same
time, one can not build the ladder out of the pair
$\hat{G}^r_{11}\bigotimes \hat{G}^r_{11}$. The reason is that the
sum over $\mathbf{k}$ of a typical ladder element, the product
$\hat{G}^{0r}_{\mathbf{k}11}\bigotimes \hat{G}^{0r}_{\mathbf{k}11}$,
where $\hat{G}^{0r}$ is given by Eq. (\ref{G0}), turns to zero,
because both functions have poles in the same semiplane of the
complex variable $E_k$. On the same reason the ladders built of
$\hat{G}^r_{22}\bigotimes \hat{G}^a_{11}$ also turn to zero, as
follows from definition (\ref{G0}). Therefore, the average $\langle
\hat{G}^r_{11}\sigma_z \hat{G}^r_{22}\sigma_z
\hat{G}^a_{11}\rangle_{\text{imp}}$ contains only one Cooperon
originating from the first two functions. Besides, the combination
$\hat{G}^r_{11}\bigotimes \hat{G}^a_{11}$ entering into this
product, results in a diffuson defined by Eq. (\ref{D}). The
corresponding Feynman diagram is shown in Fig. 2b.

In the same way, as the spin current, one may calculate the electric
charge, substituting in Eq. (\ref{JsA})  $\sigma^z k_y/2m^*$ for
$e$. This charge is given by the sum of polygons in Figs. 3a and 3b. They represent $J_{1scr}$ and $J_{2scr}$, respectively. Further, the screening electric potential is calculated within
the random phase approximation. This potential drives the spin-Hall
effect in the same way as in normal systems. \cite{Engel}

\end{document}